2023-01-14

# Semantics for a Logic of Presuppositions

*by X.Y. Newberry*

**Abstract**

In 1952 P. F. Strawson proposed a logic of presuppositions. It is an interpretation of Aristotelian logic, i.e. of the logic of the traditional syllogism. In 1981 Richard Diaz published a monograph in which he presented *truth-relevant logic.* This paper shows that truth-relevant logic is but a propositional version of the logic of presuppositions. A semantics of the logic of presuppositions is developed using truth-relevant logic. The semantics is then further extended to polyadic propositions and some consequences discussed.

# 1. Introduction

The purpose of this paper is to establish a foundation for a logic where vacuous sentences, such as $\forall x(x \neq x \rightarrow x = x+1)$, are not derivable. In such a logic Gödel's self-referential sentence is not derivable [it is vacuous], but $\sim(Ex)Prf(x, g)$ might be. Here $g$ is the Gödel number of Gödel's sentence. And $Prf(x, z)$ means that $x$ is the Gödel number of a sequence that is a proof of the sentence with Gödel number $z$. (Newberry, 2015)

In 1981 M. Richard Diaz (1981) published a monograph titled *Topics in The Logic of Relevance*. He proposed a new type of relevance logic called *truth-relevant logic*. In this logic the paradox of material implication $(P \,\&\, \sim P) \rightarrow Q$ is *not* truth-relevant, but neither are its classical equivalents such as $(P \vee \sim P) \vee Q$.

In 1952 P. F. Strawson introduced the Logic of Presuppositions (Strawson, 1952.) It conforms to Aristotle's logic of the traditional syllogism. In this logic, sentences with empty subjects are not considered true. For example the sentence "All John's children are asleep" would be considered neither true nor false if John had no children. This paper attempts to develop the semantics of the Logic of Presuppositions based on truth-relevant logic of Richard Diaz. The philosophical justification of this project can be found in Newberry (2014.)

# 2. Propositional Calculus

There are compound formulae of propositional logic such that their value can be determined by a subset of their variables. Consider for example $S = P \vee \sim P \vee Q$. When $|P|$ is $1$ then $|S| = 1$ regardless of the value of $Q$. When $|P| = 0$ then $|S| = 1$ also regardless of the value $Q$. The set of variables occurring in $S$ is $\{P, Q\}$. We say that the subset $\{P\}$ is truth-determining for $S$; for *all the valuations* of $\{P\}$, i.e. $|P| = 1$ or $|P| = 0$, we can determine the value of $S$ regardless of the other variables. (Here '$|P|$' means the Boolean value of $P$.) We use the *Boolean* values $1/0$ because we will consider as true only a proper subset of the compound formulas that evaluate as Boolean $1$. In case of atomic sentences $T = 1$, $F = 0$, in case of compound sentences, only t-relevant sentences are considered *true* when they evaluate as $1$.

In truth relevant logic the truth tables are constructed from the following elementary truth tables (Diaz, p.66.)

```
v | 1 x 0    & | 1 x 0    → | 1 x 0
----------   ----------   ----------
1 | 1 1 1    1 | 1 x 0    1 | 1 x 0
x | 1 x x    x | x x 0    x | 1 x x
0 | 1 x 0    0 | 0 0 0    0 | 1 1 1
```

Table 2.1

These tables happen to be identical with Kleene's "strong" tables. But Diaz's objective is to identify irrelevant variables, i.e. variables that do not make any contribution to the Boolean value of the compound sentence.

> **Definition 2.1:** A set of propositional variables is *truth determining* for a proposition *S* iff the truth value of *S* may be determined as *1* or *0* (using the matrices above) on all assignments of *1*, *0* to the set.

That is, it may be the case that we assign *x* to some variable, and yet we obtain a definite *1* or *0* once the entire compound formula is evaluated. It is an indication that the variable does not contribute anything to the 'output.' In this event, said variable is deemed redundaant.

Another example: $P \rightarrow (Q \rightarrow P)$ (2.1)

If $|P| = 1$ then $|(Q \rightarrow P)| = 1$ regardless of the value of *Q*, but then the entire formula (2.1) evaluates as *1* (regardless of the value of *Q*.) Suppose $|P| = 0$. Then (2.1) is true regardless of the value of *Q*. In both cases (i.e. $|P| = 1$ or $|P| = 0$) the value of (2.1) can be determined without any knowledge of the value of *Q*. Thus $\{P\}$ is truth-determining for (2.1). This is not the case for *Q*. For assume $|Q| = 1$. Then the value of $(Q \rightarrow P)$ cannot be determined without knowing the value of *P*. And without knowing the value of $(Q \rightarrow P)$, the value of (2.1) cannot be determined. (Diaz, 1981, pp. 66, 67)

> **Definition 2.2:** Let $P_1, P_2, \ldots P_n$ be all the variables occurring in *S*, $P_i$ is *truth-redundant* in *S* iff there is a truth determining set for *S* that does not contain $P_i$.

**Definition 2.3:** *S* is *truth-relevant* if it contains no truth-redundant variables.

Of special interest is the set of tautologies that are also t-relevant.

These definitions are due to Diaz (1981, pp. 65-67). We can equivalently rephrase 2.2 as follows:

**Definition 2.4:** Let $P_1, P_2, \ldots P_n$ be all the variables occurring in *S*, $P_i$ is *truth-redundant* in *S* iff the value of *S* can be determined regardless of the value of $P_i$.

# 3. Monadic Sentences

## *3.1 Definitions*

**Definition 3.1.1:** Let *S* be a first order monadic sentence. Let $P_1, P_2, \ldots P_n$ be all the atomic predicates occurring in *S*, $P_i$ is *truth-redundant* in *S* iff the truth value of *S* can be determined regardless of the interpretation of $P_{i.}$.

**Definition 3.1.2:** *S* is *truth-relevant* if it contains no truth-redundant atomic predicates.

We will use the term *satisfied* in the same sense as in classical logic. But we will say that a sentence is *true* iff it is *satisfied and t-relevant*.

The extension of the concept of truth relevance to monadic sentences is natural and straightforward. It does not present any major challenges. We will illustrate by examples.

## *3.2 "All John's children are asleep"*

Let us consider the sentence "All John's children are asleep" when John has no children (Strawson's example.) That is, consider the sentence $\forall x(Jx \rightarrow Sx)$ and an interpretation *D* such that:

|*D*| = the set of three children: Alex, Betty, Cindy; that is
|*D*| = {a, b, c}
*Jx*: *x* is John's child
*~Ja, ~Jb, ~Jc*
*Sx*: x is asleep
*Sa, Sb, ~Sc*

I.e. Alex and Betty are asleep and John has no children.

The Boolean table of *Jx* → *Sx* looks as follows:

| Row | *Jx* | *Sx* | *Jx→Sx* |
|---|---|---|---|
| 1 | 0 | 0 | **1** |
| 2 | 0 | 1 | **1** |
| 3 | 1 | 0 | **0** |
| 4 | 1 | 1 | **1** |

Table 2.1

We will now substitute constants corresponding to the elements of |*D*| one by one thus successively obtaining *Ja* → *Sa*, *Jb* → *Sb*, *Jc* → *Sc*, the corresponding Boolean values being:

| | *Jx* | *Sx* | *Jx→Sx* | Row of Table 1 |
|---|---|---|---|---|
| *Ja→Aa* | 0 | 1 | **1** | 2 |
| *Jb→Ab* | 0 | 1 | **1** | 2 |
| *Jc→Ac* | 0 | 0 | **1** | 1 |

Table 2.2

We observe that *Jx* is always *0*, and subsequently the truth value of the implication is always *1* regardless of the value of *Sx*. That is, given the interpretation *D*, the value of ∀*x(Jx* → *Sx)* can be determined without knowing the interpretation of *Sx*. And since *Sx* is not a t-relevant atomic predicate, the sentence is not t-relevant and we consider it as *~T & ~F*. It may seem contradictory that we claim that a value determination can be effected without knowing a predicate's interpretation while at the same time considering a given interpretation. But, in this case, it is the interpretation of *J* that causes the interpretation of *S* to be irrelevant.

### *3.3 Valid Sentences*

Consider ∀*x(Px* → *(Qx* → *Px))*. It is a valid sentence, and these evaluate as *1* under any interpretation. Yet even in this case there is a sense in which the interpretation of *Qx* is irrelevant. Pick an arbitrary $a$: $Pa \rightarrow (Qa \rightarrow Pa)$. Under some interpretations $|Pa| = 1$, and under some $|Pa| = 0$. It turns out that we could not care less what the interpretation of $Qa$ would or would not be. If $|Pa| = 1$ then $|(Qa \rightarrow Pa)| = 1$ regardless of the value of $Qa$, but then the entire formula evaluates as $1$ (regardless of the value of $Qa$.) Suppose $|Pa| = 0$. Then the value of the entire sentence is $1$ regardless of the value of $Qa$. In either case (i.e. $|Pa| = 1$ or $|Pa| = 0$) the value of the sentence can be determined without any knowledge of the value of $Qa$. Thus $\{Px\}$ is truth-determining for $\forall x(Px \rightarrow (Qx \rightarrow Px))$.

Let us look at prototypical cases more systematically.

### 3.4 Basic Cases

**Case 1: ∀x(Px v Qx)**

Consider $\forall x(Px \vee Qx)$, where $P$, $Q$ are atomic predicates. The formula can also be written as $\sim\exists x(\sim Px \,\&\, \sim Qx)$. Let us examine the scenarios below. The diagrams depict the extents of the (atomic) predicates $Px$, $Qx$ in some finite domain.

```
P P P P P P P P P P P P P P P P P P
Q Q Q Q . . . . . . . . . . . . . .
```

Figure 3.4.1

```
P P P P P P P P P P P P P P P P P P
. . . . . . . . . . . . . . . . . .
```

Figure 3.4.2

```
P P P P P P P P P P P P P P P P P P
Q Q Q Q Q Q Q Q Q Q Q Q Q Q Q Q Q Q
```

Figure 3.4.3

In all these cases the value of $\forall x(Px \vee Qx)$ can be determined without knowing the interpretation of $Qx$. The truth determining subset of $\forall x(Px \vee Qx)$ is $\{Px\}$. In the last scenario there are in fact two truth determining subsets {Px} and {Qx}. It is apparent that the formula will be t-relevant only if $\exists x \sim Px$ *and* $\exists x \sim Qx$. P.F. Strawson called these conditions *presuppositions*, albeit in a slightly different context.

The negation of the above formula is $\exists x(\sim Px \,\&\, \sim Qx)$. The conditions of relevance ought to be the same, i.e. $\exists x \sim Px$, $\exists x \sim Qx$, and we observe that these are satisfied automatically.

Consider $\forall x(x \leq 5 \lor x \geq 5)$. The presuppositions $\exists x(x > 5)$, $\exists x(x < 5)$ are met; the formula is t-relevant. It is also satisfied and hence true. This cannot be said for example about $\forall x(x = x \lor x \geq 5)$. The required presupposition $\exists x(x \neq x)$ is not met. The sentence is ~T & ~F.

**Case 2: ∃x(Px v Qx)**

The truth value of this sentence can be determined from $\exists x(Px)$ [or $\exists x(Qx)$.] Therefore $\exists x(Px \lor Qx)$ is never t-relevant.

The negation is $\forall x(\sim Px \,\&\, \sim Qx)$, i.e. a clean board. This sentence is always t-relevant [when satisfied.]

It follows from the aforesaid that $\boldsymbol{\forall x(Px \lor Qx) \not\models \exists x(Px \lor Qx)}$. More about this below.

It turns out that the only interesting case is the first one. Let us put $Fx =def= \sim Px$ and $Gx =def= \sim Qx$, and consider the 'existential' version $\sim\exists x(Fx \,\&\, Gx)$. The presuppositions become $\exists xFx$, $\exists xGx$. Then a typical interpretation when the sentence is satisfied and t-relevant (i.e. true) is as follows:

```
F F F F F F . . . . . . . . . . . . . . . .
. . . . . . . . . . . . . . . G G G G G G G
```

Figure 3.4.4

Under the interpretation of *F* and *G* below (Fig 3.4.5) it would be satisfied but not t-relevant (i..e. *~T & ~F*.)

```
F F F F F F . . . . . . . . . . . . . . . .      F is not relevant
. . . . . . . . . . . . . . . . . . . . . .      G
```

Figure 3.4.5

And under the next interpretation (Fig. 3.4.6) it would be false.

```
F F F F F F F F F F F F . . . . . . . . . . .
. . . . . . . . . . . G G G G G G G G G G G
```

Figure 3.4.6

Our findings are summarized in the following table.

| | Universal | Existential | Presuppositions | Negation | Presuppositions |
|---|---|---|---|---|---|
| 1 | ∀x(Px v Qx) | *~∃x(~Px & ~Qx)* | *∃x~Px & ∃x~Qx* | *∃x(~Px & ~Qx)* | automatic |
| 2 | ∀x(Px & Qx) | *~∃x(~Px v ~Qx)* | automatic | *∃x(~Px v ~Qx)* | *∃xPx & ∃xQx* |
| 3 | | *~∃x(Fx & Gx)* | *∃xFx & ∃xGx* | *∃x(Fx & Gx)* | automatic |
| 4 | | *~∃x(Fx v Gx)* | automatic | *∃x(Fx v Gx)* | *∃x~Fx & ∃x~Gx* |

Table 3.4.1

Consider for example the sentence ∀*x(Fx v Gx), Fx =def= x > 6, Gx =def= x < 8*, i.e.

$$\forall x(x > 6 \vee x < 8) \qquad (3.4.1)$$

We claim that this sentence is true (that implies t-relevant.) The presuppositions are $\exists x(x \leq 6)$ & $\exists x(x \geq 8)$, and they are both satisfied. Nevertheless no substitution instance is true. For example none of the sentences below are t-relevant.

$$5 > 6 \vee 5 < 8 \qquad (3.4.2)$$

$$7 > 6 \vee 7 < 8 \qquad (3.4.3)$$

$$9 > 6 \vee 9 < 8 \qquad (3.4.4)$$

The redundant predicates are grayed out. All the above sentences are valid but not t-relevant. The sentence (3.4.3) has two truth-determining subsets {*7 > 6*}, {*7 < 8*}.

So **$\forall x(x < 6 \vee x > 8) \nvDash (5 < 6 \vee 5 > 8)$.** The only way to "instantiate" (3.4.1) is to pick one disjunct or the other. For example when we want to substitute *5* for *x* we have to pick either *5 > 6* or *5 < 8*. (Choose wisely.)

Does this make sense? Let us briefly recapitulate how we got here. Presumably the classical definitions of *satisfaction* and *validity* do make sense. I think that my definition of an atomic predicate *relevance* also makes sense. And the result is what it is. Still, does it make sense? When some property applies to all then it must apply to one, right? Nobody is denying that this is the case with atomic predicates. Indeed $\forall x Fx \vDash \exists x Fx$. If all apples are round then there exists a round apple. But here we are talking about compound 'properties', and while $\forall x(Fx \vee Gx)$ talks about a *collection*, $Fa \vee Ga$ talks about an *individual*. You may say "All apples are green or shiny", but you would not say about any particular one "This apple is green or shiny." It makes sense to say about a collection that its members are this or that. But it does not make sense to say about one particular thing that it is this *or* that. It is one or the other or both! Just pick one and see. (Fig. 4.2.3)

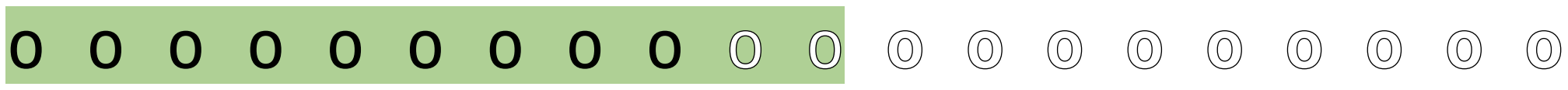

Figure 4.2.3

In t-relevant logic the *or* statement can be used only in case of uncertainty. For example you can say "It has rained or somebody has watered the lawn." You are pretty sure it is one or the other, but you do not know which one. The *or* connective cannot be used once a determination has been made or in case of necessary truths (tautologies, valid sentences, analytic sentences etc.) In particular it cannot be used about individual objects in mathematics. This is in accordance with the definitions of the logical connectives in Newberry (2019.)

$|P \vee Q| = T \;\text{ iff }\; (P + Q) \bullet \Diamond{\sim}P \bullet \Diamond{\sim}Q$

$|P \vee Q| = F \;\text{ iff }\; {\sim}(P + Q) \bullet \Diamond{\sim}P \bullet \Diamond{\sim}Q$

$|P \,\&\, Q| = T \;\text{ iff }\; P \bullet Q \bullet \Diamond P \bullet \Diamond Q$

$|P \,\&\, Q| = F \;\text{ iff }\; {\sim}(P \bullet Q) \bullet \Diamond P \bullet \Diamond Q$

where '+', '•' are a boolean operators, '◊' means *possible*.

# *4.* Polyadic Sentences

## *4.1 Definitions*

In classical logic we have

**Definition 4.1.1:**

$$(\forall x)(\forall y)Uxy \qquad (4.1.1)$$

is *true* iff for all $a_i$ **in the range of *x***

$$(\forall y)Ua_iy \qquad (4.1.2)$$

is true. What about our t-relevant logic? If we apply our classical definition "in the range of *x*" we find ourselves quantifying over sentences that are $\sim(T \vee F)$. Nothing good can possibly come out of that. In fact in our view those sentences are devoid of any meaning. The most sensible thing is to leave them out. We will therefore modify the above definition as follows:

**Definition 4.1.2:**

$$(\forall x)(\forall y)Uxy \qquad (4.1.3)$$

is *true* iff for all $a_i$ *such that* $(\forall y)Ua_iy$ *is t-relevant*

$$(\forall y)Ua_iy \qquad (4.1.4)$$

is true.

The situation is completely symmetrical with respect to *y*. The definition implies that there has to be at least one pair $<a,b>$ such that $\forall xUxb$ is true (and hence relevant) and $\forall yUay$ is true (and hence relevant.) If no such pair exists the universally quantified dyadic sentence is $\sim(T \vee F)$.

Now we need to generalize the definition of t-relevance to polyadic sentences. Hence the following definition:

> **Definition 4.1.1** A first order sentence $Q_1x_1Q_2x_2...Q_nx_nAx_1x_2 ... x_n$ with *n* variables in prenex normal form is *t-relevant* iff there is an n-tuple $<c_1, ..., c_n>$ such that
> $Q_1x_1Ax_1c_2...c_n$ is t-relevant [per definition 3.1.2], and
> $Q_2x_2Ac_1x_2...c_n$ is t-relevant, and
> . . .
> $Q_nx_nAc_1c_2...x_n$ is t-relevant.

That is to say, there has to be at least one t-relevant sentence in each ‘dimension’, and they all have to intersect at one point. In the above notation $Q_i$ means ∀ or ∃.

Now consider $\exists x \exists y Uxy$. It says that there is an *a* such that $\exists y Uay$. It means that the given $\exists y Uay$ has to be t-relevant. It implies that there is an *a* such that $\exists y Uay$ is t-relevant. But the quantifiers commute. So the same reasoning applies to $\exists y \exists x Uxy$. That is, there has to be a *b* such that $\exists x Uxb$ is t-relevant. Putting the two together we have: $\exists x \exists y Uxy$ is t-relevant iff there exists a pair <a,b> such that

> $\exists y Uay$ is t-relevant, and
> $\exists x Uxb$ is t-relevant.

By extrapolation, a first order sentence $\exists x_1 \exists x_2 \ldots \exists x_n A x_1 x_2 \ldots x_n$ with *n* variables in prenex normal form is *t-relevant* iff there is an n-tuple $<c_1, \ldots, c_n>$ such that

> $\exists x_1 A x_1 c_2 \ldots c_n$ is t-relevant [per definition 3.1.2], and
> $\exists x_2 A c_1 x_2 \ldots c_n$ is t-relevant, and
> . . .
> $\exists x_n A c_1 c_2 \ldots x_n$ is t-relevant.

We know that $\exists x \forall y Uxy$ implies $\forall x \exists y Uxy$. Therefore given a prenex sentence, we can push all the existential quantifiers inside and obtain

$$\forall x_1 \forall x_2 \ldots \forall x_m \exists x_{m+1} \exists x_{m+2} \ldots \exists x_{m+n} A x_1 x_2 \ldots x_m x_{m+1} x_{m+2} \ldots x_{m+n}$$

This will be t-relevant iff there is an n-tuple $<c_1, \ldots, c_n>$ such that

> $\forall x_1 \exists x_{m+1} \exists x_{m+2} \ldots \exists x_{m+n} A x_1 c_2 \ldots c_m x_{m+1} x_{m+2} \ldots x_{m+n}$ is t-relevant, and
> $\forall x_2 \exists x_{m+1} \exists x_{m+2} \ldots \exists x_{m+n} A c_1 x_2 \ldots c_m x_{m+1} x_{m+2} \ldots x_{m+n}$ is t-relevant, and
> . . .
> $\forall x_n \exists x_{m+1} \exists x_{m+2} \ldots \exists x_{m+n} A c_1 c_2 \ldots x_m x_{m+1} x_{m+2} \ldots x_{m+n}$ is t-relevant.

By putting it all together we obtain definition 4.1.

The reason we need an expanded definition for polyadic sentences is the following. Consider $\forall x \forall y Uxy$. It means that for all $a_i$ it is the case that $\forall y U a_i y$ [2]). But it could be that for $b \neq c$ the sentences $\forall y Uby$ and $\forall y Ucy$ are both non-relevant while having *different* redundant primitive predicates. We will illustrate by examples.

2 For the sake of brevity we assume that each object has a name, and each constant denotes.

### *4.2 Examples*

**Example 1: ~∃x∃y(Fxy & Gxy)**

```
     y
     |
     | . . . . . . . . . . . . . . . . . . . .    F or G not relevant
     | . . . . . . . . . . . . . . . . . . . .    F or G not relevant
     | . . . . . . . . . . . . . . . . . . . .    F or G not relevant
     | . . . F F F F F . . . . . . . . . . . .    F is not relevant
     | . . F F F F F F F . . . . . . . . . . .    F is not relevant
     | . . F F F F F F F F . . . . . . . . . .    F is not relevant
     | . . F F F F F F F F . . . . . . . . . .    F is not relevant
     | . . . F F F F F F . . . . . . . . . . .    F is not relevant
     | . . . . F F F F . . . . . . . . . . . .    F is not relevant
   b | . . . . . . . . . . . . . . . . . . . .    F or G not relevant
     | . . . . . . . . . . G G G G G . . . . .    G is not relevant
     | . . . . . . . . . G G G G G G G . . . .    G is not relevant
     | . . . . . . . . G G G G G G G G G . . .    G is not relevant
     | . . . . . . . . . G G G G G G G . . . .    G is not relevant
     | . . . . . . . . . . G G G G G . . . . .    G is not relevant
     | . . . . . . . . . . . . . . . . . . . .    F or G not relevant
     | . . . . . . . . . . . . . . . . . . . .    F or G not relevant
     | . . . . . . . . . . . . . . . . . . . .    F or G not relevant
     | . . . . . . . . . . . . . . . . . . . .    F or G not relevant
     -----------------------------------------------  x
                         a
```

Figure 4.2.1

The instance ~∃y*(Fay & Gay)* is t-relevant as the presuppositions ∃y*Fay* & ∃y*Gy* are satisfied (Table 3.4.1, line 3.) The instance ~∃x*(Fxb & Gxb)* is *not*. Therefore there is *no* pair <a,b> such that both ~∃y*(Fay & Gay)* and ~∃x*(Fxb & Gxb)* are t-relevant.

**Example 2: ~∃x∃y(Fxy & Gxy)**

Here is a different example.: *~∃x∃y(Fxy & Gxy) = ∀x∀y(~Fxy v ~Gxy),*
*Fxy =def= x + y < 8, Gxy =def= x + y > 8,*
i.e. *~∃x∃y(x + y < 8 & x + y > 8) = ∀x∀y(x + y >= 8 v x + y <= 8).*

```
         y
         |
17       | G G G G G G G G G G G G G G G G G G G G
16       | G G G G G G G G G G G G G G G G G G G G
15       | G G G G G G G G G G G G G G G G G G G G
14       | G G G G G G G G G G G G G G G G G G G G
13       | G G G G G G G G G G G G G G G G G G G G
12       | G G G G G G G G G G G G G G G G G G G G
11       | G G G G G G G G G G G G G G G G G G G G
10       | G G G G G G G G G G G G G G G G G G G G
9        | G G G G G G G G G G G G G G G G G G G G
8        | . G G G G G G G G G G G G G G G G G G G
7        | F . G G G G G G G G G G G G G G G G G G
6        | F F . G G G G G G G G G G G G G G G G G
5        | F F F . G G G G G G G G G G G G G G G G
4        | F F F F . G G G G G G G G G G G G G G G
3        | F F F F F . G G G G G G G G G G G G G G
2        | F F F F F F . G G G G G G G G G G G G G
1        | F F F F F F F . G G G G G G G G G G G G
         -----------------------------------------------  x
                             1 1 1 1 1 1 1 1 1 1 2
           1 2 3 4 5 6 7 8 9 0 1 2 3 4 5 6 7 8 9 0
```

Figure 4.2.2

In this scenario the formula is t-relevant. Not all substitution instances are t-relevant. By substituting any number from the interval [1,7] for *x* we obtain a t-relevant formula. For example substitute *3*:

∀y(**3** + y >= 8 v **3** + y <= 8)

The sentence is t-relevant as the presuppositions ∃y*~(3 + y >= 8)* and ∃y*~(3 + y <= 8)* are both satisfied. Substitute 13:

∀y(**13** + y >= 8 v **13** + y <= 8)

This sentence is not t-relevant as the presuppositions ∃y~*(13 + y >= 8) i.e.* ***∃y(13 + y < 8)*** is *not* satisfied. Note that no substitution instance of both *x* and *y* is t-relevant. E.g.

6 + 7 >= 8 v 6 + 7 <= 8

The presuppositions are **6 + 7 < 8** & 6 + 7 > 8. There is no chance the former could be satisfied. There are no two natural numbers such that their sum is simultaneously less and greater than 8.

Most of the sentences ~∃*y(Fay & Gay)* in the diagram are not t-relevant, hence *~T & ~F*. We only want to quantify over the t-relevant sentences, i.e. when *b = [1,7]*. But according to figure 4.2.2 we are doing that, and we can simply disregard that we are also quantifying over the non-relevant ones. Nevertheless the question arises if it could be that all the t-relevant instances were satisfied while ~∃*x*∃*y(Fxy & Gxy)* was not. The answer is no because any counterexamples to ~∃*x*∃*y(Fxy & Gxy)* must necessarily be t-relevant.

### Theorem 4.2.1

Let ~∃*x*∃*y(Fxy & Gxy)* be a sentence with atomic predicates *F, G*. If all the t-relevant instances ~∃*y*($Fa_iy$ & $Ga_iy$) and ~∃*x*($Fxb_i$ & $Gxb_i$) are satisfied then ~∃*x*∃*y(Fxy & Gxy* is satisfied. (*Satisfied* means *evaluates as Boolean 1*.)

**Proof:** Assume the opposite. Then ∃*x*∃*y(Fxy & Gxy)*. So there is some *<c,d>*, say *<11,12>* (Fig. 4.2.2), such that

∃*y(Fcy & Gcy)* and ∃*x(Fxd & Gxd)* (4.2.1)

Both of these sentences are t-relevant, and so are their negations

~∃*y(Fcy & Gcy)* and ~∃*x(Fxd & Gxd)* (4.2.2)

These are t-relevant instances of ~∃*x*∃*y(Fxy & Gxy)*. If such *<c,d>* do exist then (4.2.2) are t-relevant albeit false. But this contradicts the assumption that all the t-relevant instances are satisfied. QED.

**Example 3: ~∃x∃y(Lxy & My)**

Let *Lxy =def= (x + y <= 8), My =def= (y = 12)*. We obtain *~∃x∃y(x + y <= 8 & y = 12)*. The above formula is not t-relevant according to Figure 4.4.1 below.

```
        y
        |
17      | . . . . . . . . . . . . . . . . . . . .
16      | . . . . . . . . . . . . . . . . . . . .
15      | . . . . . . . . . . . . . . . . . . . .
14      | . . . . . . . . . . . . . . . . . . . .
13      | . . . . . . . . . . . . . . . . . . . .
12      | M M M M M M M M M M M M M M M M M M M M
11      | . . . . . . . . . . . . . . . . . . . .
10      | . . . . . . . . . . . . . . . . . . . .
9       | . . . . . . . . . . . . . . . . . . . .
8       | L . . . . . . . . . . . . . . . . . . .
7       | L L . . . . . . . . . . . . . . . . . .
6       | L L L . . . . . . . . . . . . . . . . .
5       | L L L L . . . . . . . . . . . . . . . .
4       | L L L L L . . . . . . . . . . . . . . .
3       | L L L L L L . . . . . . . . . . . . . .
2       | L L L L L L L . . . . . . . . . . . . .
1       | L L L L L L L L . . . . . . . . . . . .
        ----------------------------------------------- x
                          1 1 1 1 1 1 1 1 1 1 2
          1 2 3 4 5 6 7 8 9 0 1 2 3 4 5 6 7 8 9 0
```

Figure 4.4.1

The sentence *~∃x∃y(x + y <= 8 & y = 12)* bears a strong resemblance to Gödel’s sentence. Let

$$\sim(\exists x)(\exists z)(Prf(x, z) \ \& \ Diag(y, z)) \qquad \text{(U)}$$

be a formula with one free variable *y*,, where *Diag(y, z)* means that *y* is the Gödel number of a formula with one free variable iff *z* is the Gödel number number of the formula obtained from *y* by substituting (the numeral of) the Gödel number of *y* for the free variable in *y*. And *Prf(x, z)* means that *x* is the Gödel number of a sequence that is a proof of the sentence with Gödel number *z*.

Let the constant *k* be the Gödel number of (U). We substitute *k* for the free variable *y* in (U), and obtain

$$\sim(\exists x)(\exists z)(Prf(x,z) \,\&\, Diag(k,z)) \qquad \text{(G)}$$

This is Gödel's self-referential sentence. Then graphically the situation will look much like Figure 4.4.1 in the sense that the pairs $<x,y>$ of natural numbers picked by *Prf()* and *Diag()* respectively will not intersect. That means that *G* will *not* be t-relevant even if $\sim\exists x(Prf(x, \ulcorner G \urcorner)$ is true and provable!

The sentence G is ~T & ~F in the logic of presuppositions since $\sim(\exists x)Prf(x, \ulcorner G \urcorner)$ contradicts the *presupposition* $(\exists x)Prf(x, \ulcorner G \urcorner)$. This renders *Diag(k,z)* redundant. However *Prf* and *Diag* are compound predicates while in t-relevant logic the relevance attribute applies only to atomic predicates.

Let us take a closer look. Note that $\ulcorner U(y) \urcorner$, is a specific natural number. When we expand $Diag(\ulcorner U(y) \urcorner, z)$ then $\ulcorner U(y) \urcorner$, will appear in some predicate $v = \ulcorner U(y) \urcorner$ for some variable *v*. This atomic predicate will be redundant.

# Bibliography

Diaz, M.R. (1981) *Topics in the Logic of Relevance*, Munich, Germany: Philosophia Verlag.

Newberry, X.Y. (2014) Meaning, Presuppositions, Truth-relevance, Gödel's Sentence and the Liar Paradox, https://philpapers.org/archive/NEWMPT.pdf

Newberry, X.Y. (2015) The Diagonal Lemma Fails in Aristotelian Logic, *The Reasoner* Volume 10, Number 1, http://arxiv.org/abs/1510.0439

Newberry, X.Y. (2019) Truth-relevant Logic – Propositional Calculus, https://arxiv.org/abs/2204.10887

Strawson, P.F. (1952) *Introduction to Logical Theory*, London: Methuen.